\documentclass[usenatbib]{mn2e}
\usepackage{graphicx}

\def\aj{AJ}                   
\def\apjl{ApJ}                
\def\aap{A\&A}                
\def\mnras{MNRAS}             
\def\nat{Nature}              

\begin{document}
\label{start}

\title[The Capture of Centaurs as Trojans]{The Capture of Centaurs as Trojans}
\author[Horner \& Evans]{J. Horner$^1$ and N. Wyn Evans$^2$ \\
$^1$ Physikalisches Institut, University of Bern, Sidlerstrasse
5, 3012 Bern, Switzerland \\
$^2$ Institute of Astronomy, University of Cambridge,
Madingley Road, Cambridge CB3 0HA, UK}
\date{Received / Accepted}

\maketitle

\begin{abstract}
Large scale simulations of Centaurs have yielded vast amounts of data,
the analysis of which allows interesting but uncommon scenarios to be
studied. One such rare phenomenon is the temporary capture of Centaurs
as Trojans of the giant planets. Such captures are generally short (10
kyr to 100 kyr), but occur with sufficient frequency ($\sim 40$
objects larger than 1 km in diameter every Myr) that they may well
contribute to the present-day populations.  Uranus and Neptune seem to
have great difficulty capturing Centaurs into the 1:1 resonance, while
Jupiter captures some, and Saturn the most ($\sim 80 \%$). We
conjecture that such temporary capture from the Centaur population may
be the dominant delivery route into the Saturnian Trojans. Photometric
studies of the Jovian Trojans may reveal outliers with Centaur-like as
opposed to asteroidal characteristics, and these would be prime
candidates for captured Centaurs.
\end{abstract}


\section{Introduction}
Lagrange was the first to observe that there is an exact solution of
the three body problem in which the bodies lie at the vertices of an
equilateral triangle. This has a direct application to the Solar
system. The Trojan asteroids librate about the so-called $L_4$ and
$L_5$ Lagrange points, and lie roughly $60^\circ$ ahead
and behind the mean longitude of the planet.  (e.g., Danby 1988).

Jupiter provides the best known and longest studied case.  The
population of the Jovian Trojans is substantial. For example, the
number of objects with radii in excess of 1 km may exceed $\sim 10^5$
in total (Jewitt, Trujillo \& Luu 2000). By contrast, very few Trojans
of the other planets are known. Only 3 Mars Trojans (namely (5261)
Eureka, 1998 VF$_{31}$ and 1999 UJ$_7$) have been securely identified
(see e.g., Tabachnik \& Evans 1999; Rivkin et al. 2003). Recent
wide-field surveys of the outer Solar system (Chiang et al. 2003;
Sheppard \& Trujillo 2005) have also discovered 2 Neptunian Trojans
(namely 2001 QR$_{322}$ and 2004 UP$_{10}$). There have been surveys
for Trojans of Saturn, Uranus, and the Earth, but they have not
yielded any positive detections (e.g., Whiteley \& Tholen 1998;
Sheppard \& Trujillo 2005).  Nonetheless, numerical simulations by a
number of authors (e.g., Holman \& Wisdom 1987; Evans \& Tabachnik
2000) suggest that Trojans could exist in long-lived and stable orbits
in the vicinity of these planets.

A number of possible formation scenarios for the Jovian Trojan
asteroids have been proposed. One suggestion is that the Trojans are
planetesimals formed near, and captured by, the growing Jupiter
possibly with the aid of a dissipative mechanism like gas drag or
collisions (e.g., Marzari \& Scholl 1998; Fleming \& Hamilton 2000).
Another possibility is that the Trojans were captured into co-orbital
motion with Jupiter in the early Solar system, during the time of
migration of the giant planets (Morbidelli et al. 2005). The origin of
the Trojan asteroids of Mars and Neptune may however be different to
the Jovian case. Chiang et al. (2003) suggest that debris from
planetesimal collisions occurring after Neptune reached its current
location may have accreted naturally in the $1:1$ resonance to provide
its Trojan clouds.

In this {\it Letter}, we consider the possibility that some of the
Trojans may originate from capture of Centaurs. This idea seems to
have been first suggested by Rabe (1970), but hard evidence from
numerical simulations has so far been lacking.  The mechanism of
capture is often invoked to explain the irregular outer satellites of
Jupiter, Uranus and Neptune (e.g., Sheppard \& Jewitt 2003; Sheppard
et al. 2005). Here, we supply examples from our suite of numerical
integrations to confirm that it can also provide Trojans.

\begin{table}
\begin{center}
\begin{tabular}{c|c|c}\hline\hline Object Type & Number & $\langle
  T_{\rm cap} \rangle $ (in kyr)\\
\hline\hline
Jupiter Trojans  &  10 &  81 \\  
Saturn Trojans   &  54 &  37 \\
Uranus Trojans   &   3 & 139 \\
Neptune Trojans  &   0 & - \\
\hline
Jovian Satellites      & 1  & -  \\
Saturnian Satellites   & 0  & -  \\
Uranian Satellites     & 0  & -  \\
Neptunian Satellites   & 0  & -  \\ \hline\hline
\end{tabular}
\end{center}
\caption{The numbers of objects captured as Trojans or satellites of
the giant planets during 3 Myr integrations of 23\,328 Centaur-like
objects. $\langle T_{\rm cap} \rangle$ gives the mean duration of
these capture events.}
\label{tab:capture}
\end{table}

\section{Simulations}

In order to understand the behaviour of the comet-like Centaurs, 32 of
them were chosen as the subject for large-scale numerical
integrations.  The orbits of each of the chosen objects -- as given by
{\it The Minor Planet Center} in June 2002 -- were then incrementally
modified to give 729 "clones", which formed a $9 \times 9 \times 9$
grid in the space of semimajor axis $a$, eccentricity $e$ and
inclination $i$.  This gave a total of 23\,328 test particles, which
were then followed for 3 Myr under the gravitational influence of the
Sun, Jupiter, Saturn, Uranus and Neptune, using the {\tt MERCURY}
integrator (Chambers 1999). The gravitational effects of the
terrestrial planets were neglected. A more detailed exposition of the
simulations is given in Horner et al. (2004a,b).

One of the areas of interest that prompted the integrations was the
question of the temporary capture of objects by the giant outer
planets, both to Trojan-like and satellite-like orbits. The orbits of
the clones were recorded only at 100 year intervals. Although
short-lived satellite-like behaviour, such as that displayed by comets
P/Helin-Roman-Crockett (Tancredi et al. 1990) and P/Gehrels 3 (Rickman
et al. 1981), is missed, longer-term captures are detectable within
our numerical dataset. Consequently, the data were searched for
objects whose semimajor axis stayed within one Hill radius of that of
Jupiter, Saturn, Uranus and Neptune -- temporarily captured as either
a moon or a Trojan. These provisional candidates were then examined in
the co-rotating frame of the planet's orbit to see whether they had
Trojan-like or satellite-like behaviour. To ensure that an object was
indeed captured, a minimum number of 800 orbital periods was
required. For example, in the case of Jupiter, the object had to show
Trojan-like or satellite-like behaviour for at least 9.5 kyr ($\sim
800$ orbital periods of Jupiter).

\begin{figure}
\begin{center}
\includegraphics[width=0.7\hsize]{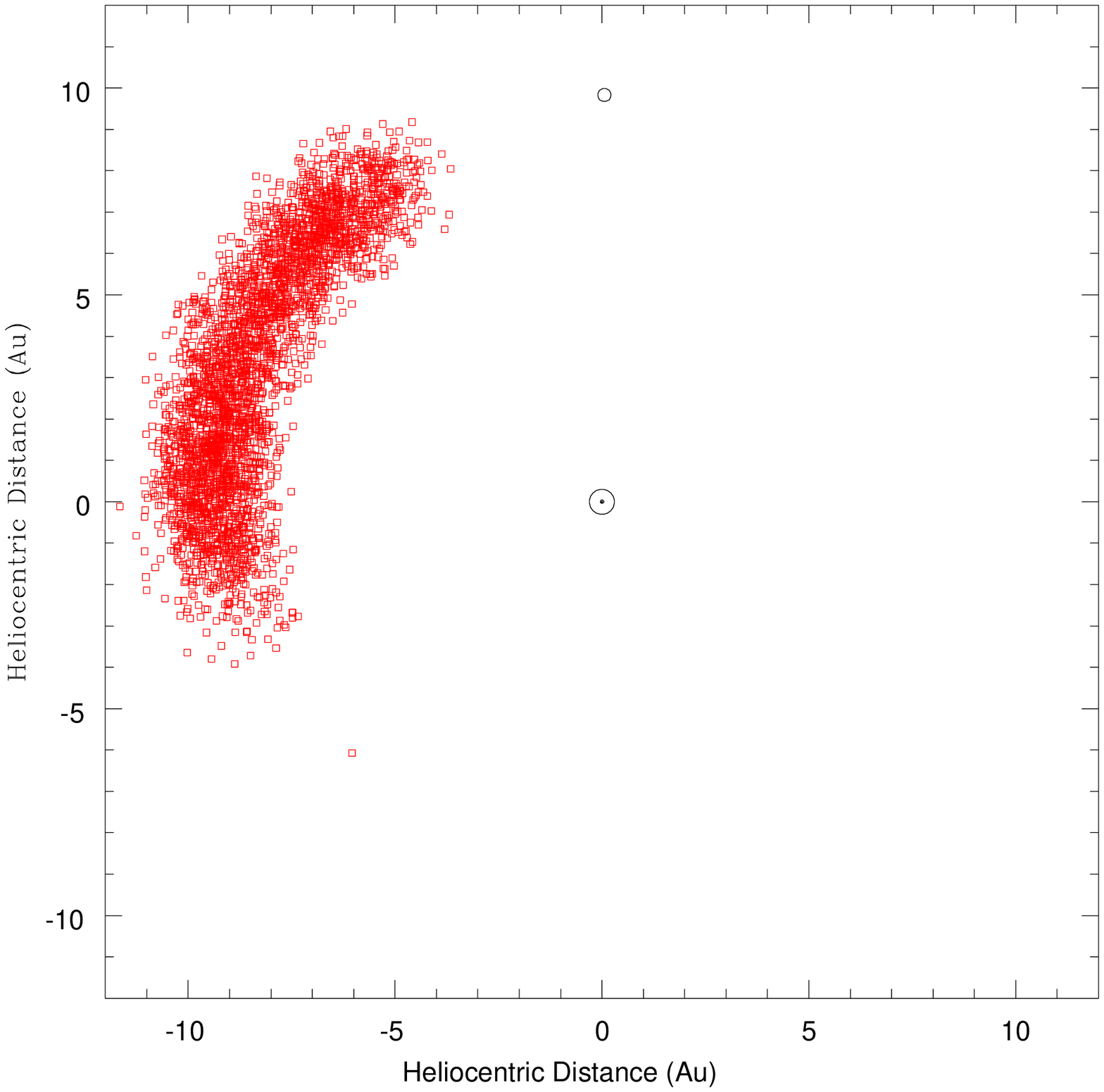}
\end{center}
\includegraphics[width=\hsize]{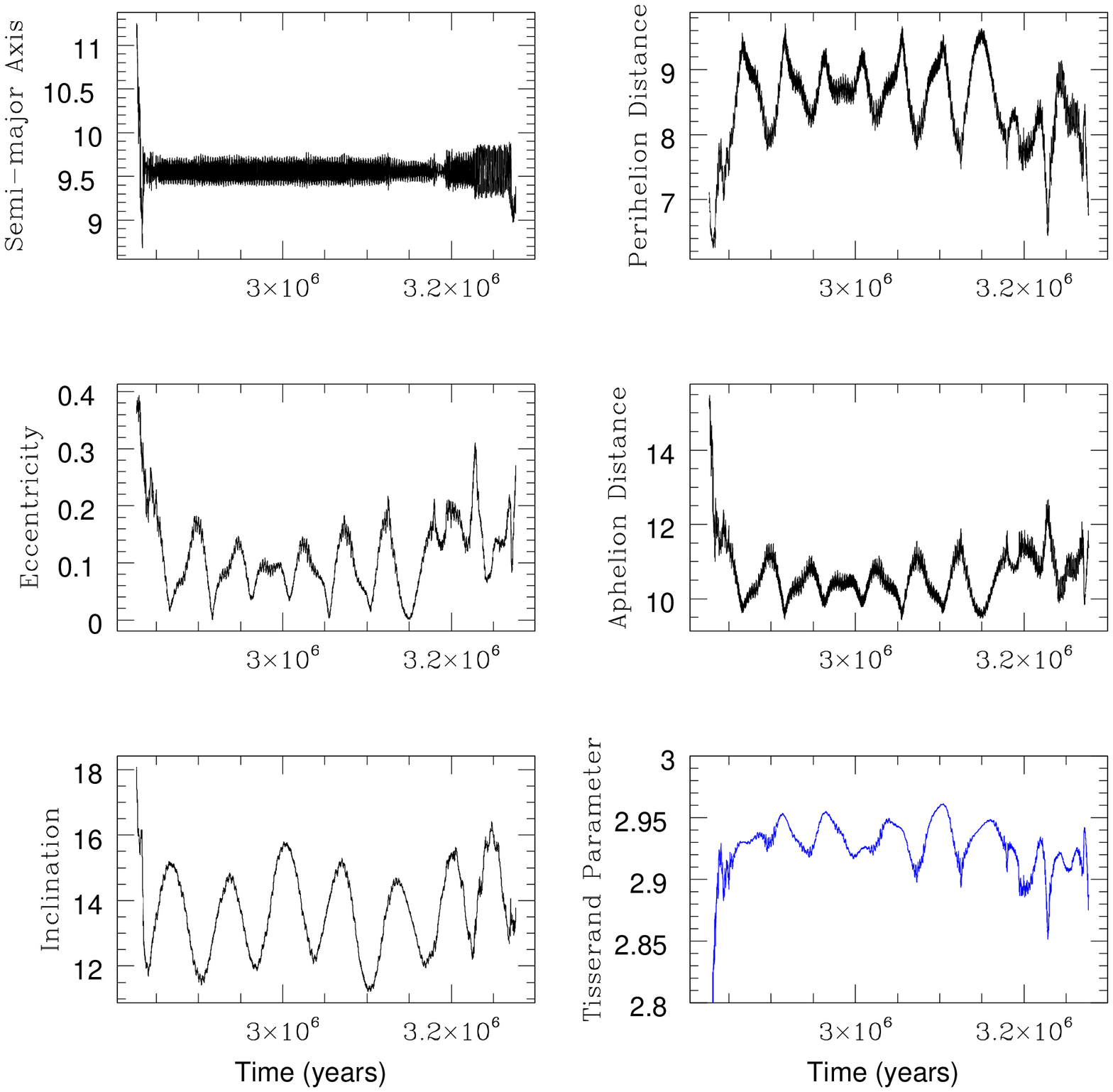}
\caption{\label{fig:sat} Capture of a clone of (10199) Chariklo as a
Saturnian Trojan. The upper panel shows the orbit plotted in a frame
co-rotating with Saturn. The positions of the Sun and Saturn are
marked. The lower panel shows the evolution of the orbital properties.
The Tisserand parameter $T_{\rm S}$ is evaluated with respect to
Saturn.  [The initial semimajor axis, eccentricity and inclination of
the clone are $a = 15.724$ au, $e = 0.154$ and $i = 23.46^\circ$.]}
\end{figure}
\begin{figure}
\begin{center}
\includegraphics[width=0.7\hsize]{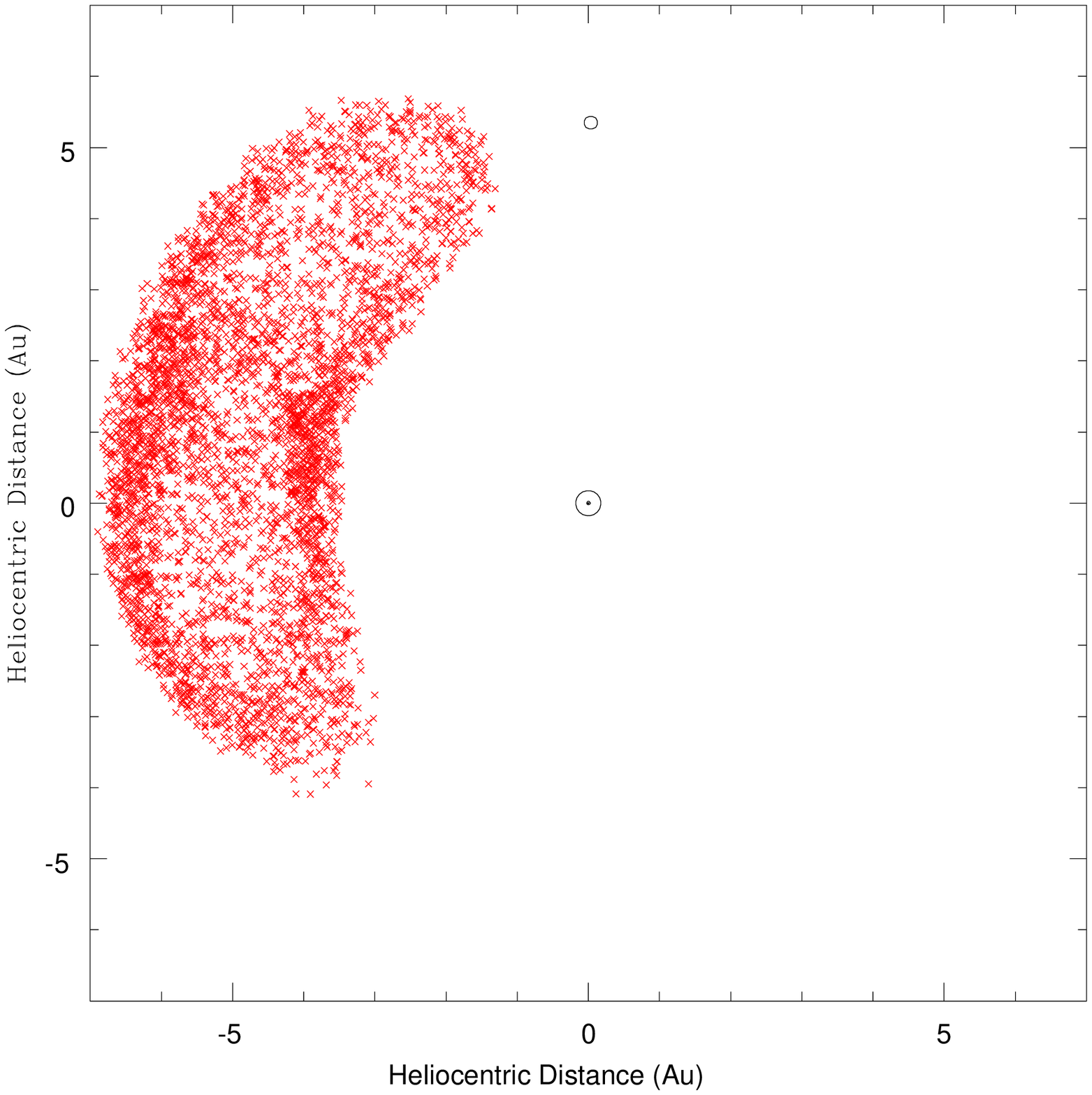}
\end{center}
\includegraphics[width=\hsize]{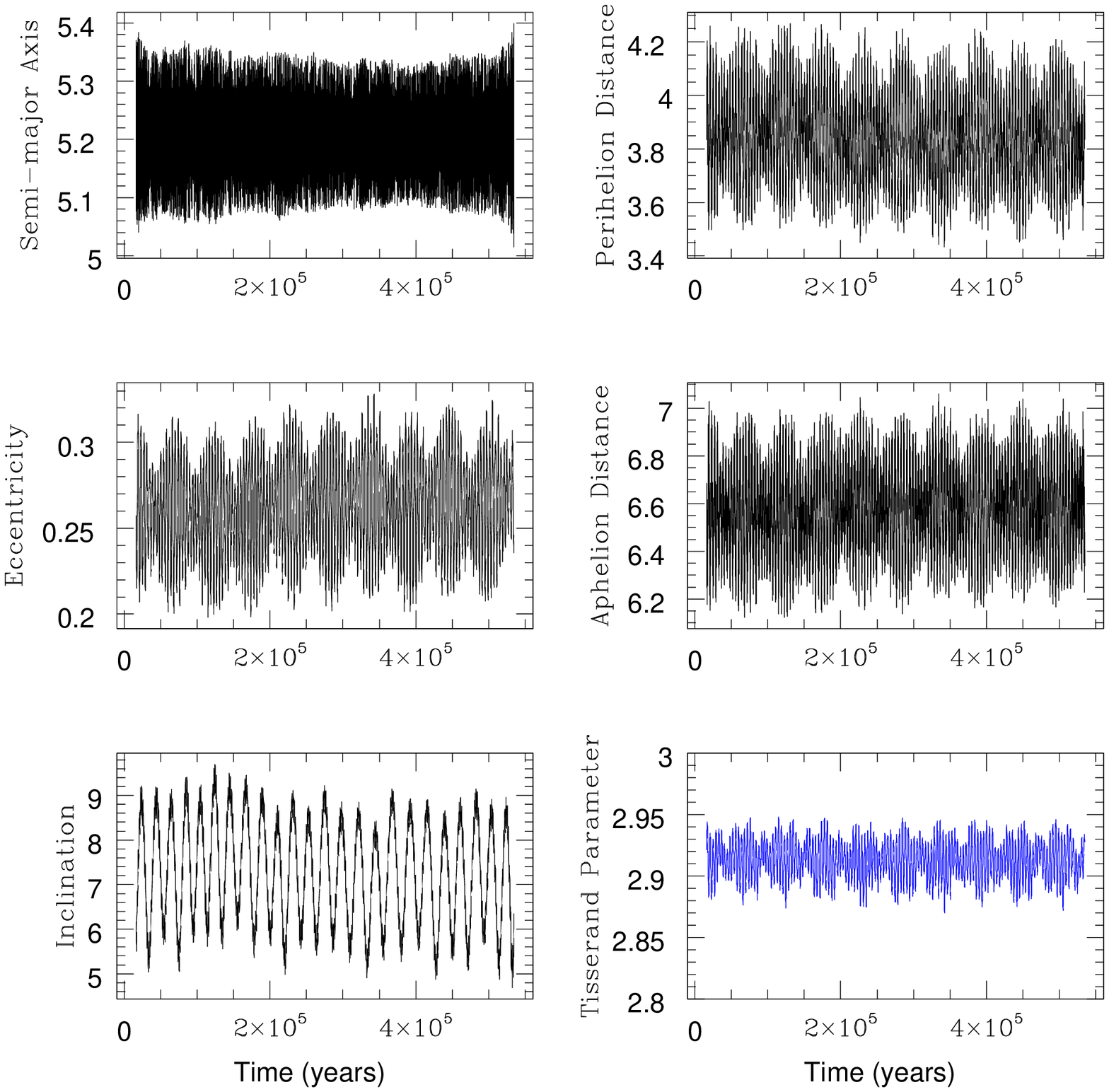}
\caption{\label{fig:jove} Capture of a clone of 1996 AR$_{20}$ as a Jovian
Trojan. The upper panel shows the orbit plotted in a frame co-rotating
with Jupiter. The lower panel shows the evolution of semimajor axis,
perihelion and aphelion distance (all in au), inclination (in degrees)
and eccentricity and the Tisserand parameter with respect to Jupiter
$T_{\rm J}$.  [The initial semimajor axis, eccentricity and
inclination of the clone are $a = 15.177$ au, $e = 0.617$ and $i =
6.17^\circ$.]}
\end{figure}

\section{Results and Discussion}

From a total sample of 23\,328 objects, 67 were captured as Trojans
for a period of 800 or more orbital periods. This is approximately
$0.3\%$ of the sample. The numbers are broken down according to each
giant planet in Table~\ref{tab:capture}.  It is interesting that --
even though there are no known Saturnian Trojans -- Saturn is more
efficient at capturing Centaurs temporarily into Trojan-like orbits
than Jupiter. Table~\ref{tab:capture} also gives the mean duration
$\langle T_{\rm cap} \rangle$ of the capture events. Of course,
Newton's equations of motion are time-reversible, so some form of
dissipation is required for permanent capture. All the captures in our
dataset are temporary with average durations of the order of kyrs. It
must be remembered that no capture events shorter than 800 orbital
periods were examined and that such short captures could be the most
common type of event.

Holman \& Wisdom (1992) carried out surveys of the stability of test
particles placed in the vicinity of the Lagrange points of Jupiter and
Saturn. They noted that the stable regions are much more ragged for
Saturn than for Jupiter. In the case of Saturn, the stability zone is
disrupted by islands of instability, possibly caused by the near $5:2$
resonance between Jupiter and Saturn (Innanen \& Mikkola
1989). Therefore, it is understandable that the long-lived population
of Jovian Trojans is larger than that of Saturn. Our calculations
raise the possibility that the reverse may pertain to temporary
captures. Such temporary Saturnian Trojans may well exist in
substantial numbers. It may be that for Saturn the main delivery
mechanism of Trojans is temporary capture from the Centaur region,
rather than primordial capture of planetesimals.

Let us give two examples out of the 67 events listed in
Table~\ref{tab:capture}. Figure~\ref{fig:sat} shows the temporary
capture of a clone of (10199) Chariklo into the $1:1$ resonance with
Saturn. The upper panel shows the projection of the orbit onto the
invariable plane in a frame co-rotating with Saturn. It is clear that
the clone follows a tadpole orbit (e.g., Murray \& Dermott 2000),
librating about the leading or $L_4$ Lagrange point.  The temporary
Trojan phase lasts for $\sim 400$ kyr, during which the eccentricity
$e$ and inclination $i$ librations are modest compared to its prior
and subsequent evolution.

Figure~\ref{fig:jove} shows the temporary capture of a clone of 1996
AR$_{20}$ into the $1:1$ resonance with Jupiter.  The upper panel again
shows that the clone is librating about the $L_4$ Lagrange point.  The
clone displays moderate variations in semimajor axis $a$, eccentricity
$e$ and inclination $i$ whilst in the resonance. It resides in the
resonance for $\sim 0.5$ Myr before ejection from the Solar
system. This is the longest example of Trojan capture in our dataset.

In addition to the Trojan-like objects, we also searched for temporary
satellite-like captures of Centaurs. As shown in
Table~\ref{tab:capture}, this is a much rarer occurrence. We found
only one convincing example -- a clone of (32532) Thereus (or 2001
PT$_{13}$) displayed behaviour which hints at a temporary moon capture by
Jupiter. Although it has often been conjectured that the outer
irregular satellites of the giant planets may have been captured, this
seems to be a scarcer phenomenon that Trojan capture in our dataset.

In previous work (Horner et al. 2004a), the current population of the
Centaurs with nuclei larger than $\sim 1$ km in diameter was estimated
at $\sim 44300$. Using the results of our simulations, we reckon that
the capture rate of Centaurs as temporary Trojans is $\sim 40$ objects
every Myr (for lengthy captures). The average duration of such a
capture is a few tens of thousands of years. Given the length of time
the clones can spend in these stable orbits, it is quite possible that
there are objects lurking in the Trojan clouds of the outer planets
which are temporary Centaur captures. In fact, simulations of the
Jovian Trojans show that at least two objects currently classified as
Trojans are experiencing only a brief visit to the region, rather than
a prolonged stay (Karlsson 2004, and simulations by the authors).

The permanent Trojans of Jupiter are known to be more populous in the
leading ($L_4$) cloud than in the trailing ($L_5$) cloud. Analyzing
the captured objects in our dataset, no such trend is
present. Table~\ref{tab:L4-5populations} shows the breakdown of the
capture locations for our sample. It seems that the captured
population has a different profile to the permanent one, with nearly
equal likelihood of capture in the L4 or L5 regions. Finally,
Table~\ref{tab:L4-5populations} also records the fact that few of the
captures are into horseshoe orbits. One possible reason is that such
orbits are significantly less stable than their tadpole
brethren. Hence, captures in such orbits may be far less likely to
survive the 800 revolutions required for detection in this survey.

\begin{table}
\begin{center}
\begin{tabular}{c||c|c|c}\hline\hline Planet & $L_4$ capture & $L_5$ capture & 
Horseshoe \\
\hline\hline
Jupiter Trojans  &   2   &   5  &   3   \\  
Saturn Trojans   &  25   &  26  &   3   \\
Uranus Trojans   &   2.5 &   0  &   0.5 \\
Neptune Trojans  &     0 &   0  &   0   \\
\hline \hline
\end{tabular}
\end{center}
\caption{The number of objects captured into the leading (L4) and
trailing (L5) Trojan clouds, along with those objects captured onto
horseshoe orbits. Objects that displayed two periods of Trojan
behaviour in different regions gave a score of 0.5 in each region
occupied.}
\label{tab:L4-5populations}
\end{table}

\section{Conclusions}

This {\it Letter} has demonstrated a new possibility for the origin of
some of the Trojans of the giant planets. They may be Centaurs,
temporarily captured into the $1:1$ resonances. We have used data from
3 Myr simulations of representatives of the Centaur population to
provide specific examples of this delivery mechanism. In particular,
Saturn seems to be most efficient at making such temporary captures
from the Centaur region. Saturn captures the bulk of the $\sim 40$
Centaurs every Myr that pass through a lengthy temporary Trojan phase
(a capture for 800 or more orbital periods of the parent
object). Since we expect $\sim 40$ such lengthy captures, it is quite
likely that the number of shorter captures is significantly higher,
and hence that there may well be such temporary visitors residing in
these regions at the present time. The objects captured within these
simulations display a roughly equal likelihood of capture into the
leading and trailing Trojan clouds, a quantitative difference to the
observed long-lived population of Jovian Trojans.

Possible evidence of such interlopers within the Trojan clouds might
be garnered from observations of colours or from photometric
activity. For example, it would be interesting to see whether any
Jovian Trojans display cometary out-gassing, since recently captured
Centaurs may still contain volatiles, whilst any Trojans captured
since the birth of the Solar system are unlikely to display such
activity. Similarly, if any Jovian Trojans are found be of
significantly different colour to other objects in the cloud, this may
hint at a different delivery mechanism, and may help the
identification of such temporary visitors. At Saturn, since it is
unlikely that many Trojans would survive at the $L_4$ and $L_5$ points
on timescales approaching the age of the Solar system, it is likely
that any Trojans discovered in the future represent recent, temporary
captures, rather than a native population.

Morbidelli et al. (2005) suggested that the Jovian Trojan population
may have been captured into co-orbital motion with Jupiter during the
latter part of its proposed migration. The fact that temporary
captures can be seen in our dataset with such frequency seems to add
weight to this mechanism for the formation of the Trojan clouds, by
illustrating that temporary captures are not uncommon even at the
current day. Given the fact of Jupiter's migration, it is quite
feasible that objects originally captured on temporary orbits could be
converted to ones which could reside in the Trojan region for the age
of the Solar System.

\label{finish}

\begin{thebibliography}{}

\bibitem[Chambers(1999)]{1999MNRAS.304..793C} Chambers J.~E.\ 1999,
\mnras, 304, 793

\bibitem[Chiang et al.(2003)]{2003AJ....126..430C} Chiang E.~I., et al.\ 
2003, \aj, 126, 430 

\bibitem[Danby 1988]{danby} Danby J.M.A. 1988, Fundamentals of
Celestial Mechanics (Willmann-Bell, Richmond)

\bibitem[Evans \& Tabachnik(2000)]{2000MNRAS.319...80E} Evans N.~W., 
Tabachnik S.~A.\ 2000, \mnras, 319, 80  

\bibitem[Fleming \& Hamilton(2000)]{2000Icar..148..479F} Fleming H.~J.,  
Hamilton D.~P.\ 2000, Icarus, 148, 479 

\bibitem[Holman \& Wisdom 1993]{wh93} Holman M.J., Wisdom J. 1993, AJ,
105, 1987

\bibitem[Horner et al.(2004)]{2004MNRAS.354..798H} Horner J., Evans 
N.~W., Bailey M.~E.\ 2004a, \mnras, 354, 798 

\bibitem[Horner et al.(2004)]{2004MNRAS.355..321H} Horner J., Evans 
N.~W., Bailey M.~E.\ 2004b, \mnras, 355, 321 

\bibitem[Innanen \& Mikkola (1989)]{im}
Innanen K., Mikkola S., 1989, AJ, 97, 900

\bibitem[Jewitt et al.(2000)]{2000AJ....120.1140J} Jewitt D.~C., Trujillo 
C.~A., Luu J.~X.\ 2000, \aj, 120, 1140 
  
\bibitem[Karlsson(2004)]{2004A&A...413.1153K} Karlsson O.\ 2004, \aap, 
413, 1153 
 
\bibitem[Marzari \& Scholl(1998)]{1998Icar..131...41M} Marzari F., 
Scholl H. 1998, Icarus, 131, 41  

\bibitem[Morbidelli et al.(2005)]{2005Natur.435..462M} Morbidelli A., 
Levison H.~F., Tsiganis K., Gomes R.\ 2005, \nat, 435, 462 
 
\bibitem[Murray \& Dermott(2000)]{Mu00} Murray C.~D., 
Dermott S.~F. 2000, Solar System Dynamics, Cambridge University Press, 
sec. 3.9

\bibitem[Rabe(1972)]{1972IAUS...45...55R} Rabe, E.\ 1972, IAU Symp.~ 45: 
The Motion, Evolution of Orbits, and Origin of Comets, p 55 
 
\bibitem[Rickman \& Malmort(1981)]{1981A&A...102..165R} Rickman H.,
Malmort A.~M.\ 1981, \aap, 102, 165

\bibitem[Rivkin et al.(2003)]{2003Icar..165..349R} Rivkin A.~S., Binzel 
R.~P., Howell E.~S., Bus S.~J., Grier J.~A.\ 2003, Icarus, 165, 349 

\bibitem[Sheppard \& Jewitt(2003)]{2003Natur.423..261S} Sheppard S.~S.,
Jewitt, D.~C.\ 2003, \nat, 423, 261 

\bibitem[Sheppard \& Trujillo(2005)]{2005DPS....37.5615S} Sheppard S.~S., 
Trujillo C.\ 2005, AAS/Division for Planetary Sciences Meeting 
Abstracts, 37.5615

\bibitem[Sheppard et al.(2005)]{2005AJ....129..518S} Sheppard S.~S., 
Jewitt D., Kleyna J.\ 2005, \aj, 129, 518 
 
\bibitem[Tancredi et al.(1990)]{1990A&A...239..375T} Tancredi G., 
Lindgren M., Rickman H.\ 1990, \aap, 239, 375 

\bibitem[Tabachnik \& Evans(1999)]{1999ApJ...517L..63T} Tabachnik S., 
Evans N.~W.\ 1999, \apjl, 517, L63 

\bibitem[Whiteley \& Tholen 1998]{wt} Whiteley R.J., Tholen D.J. 1998,
Icarus, 136, 154
  
\end{thebibliography}
\end{document}